\documentclass[pdflatex,sn-mathphys-num]{sn-jnl}

\usepackage{pdfpages}
\usepackage{graphicx}%
\usepackage{multirow}%
\usepackage{amsmath,amssymb,amsfonts}%
\usepackage{amsthm}%
\usepackage{mathrsfs}%
\usepackage[title]{appendix}%
\usepackage{xcolor}%
\usepackage{textcomp}%
\usepackage{manyfoot}%
\usepackage{booktabs}%
\usepackage{algorithm}%
\usepackage{algorithmicx}%
\usepackage{algpseudocode}%
\usepackage{listings}%

\begin{document}
\title[Article Title]{Scalable Lithium Niobate Nanoimprinting for Nonlinear Metalenses}
\author*[1]{\fnm{Ülle-Linda} \sur{Talts}}\email{utalts@ethz.ch}
\equalcont{These authors contributed equally to this work.}
\author[1]{\fnm{Helena} \sur{Weigand}}\email{hweigand@ethz.ch}
\equalcont{These authors contributed equally to this work.}

\author[1]{\fnm{Irene} \sur{Occhiodori}}\email{iocchiodori@student.ethz.ch}

\author*[1]{\fnm{Rachel} \sur{Grange}}\email{grangera@ethz.ch}

\affil[1]{\orgname{ETH Zurich}, \orgdiv{Department of Physics, Institute for Quantum Electronics, Optical Nanomaterial Group}, \street{Auguste-Piccard-Hof, 1}, \postcode{8093}, \city{Zurich}, \country{Switzerland}}


\maketitle
Miniaturizing nonlinear optical components is essential for integrating advanced light manipulation into compact photonic devices, enabling scalable and cost-effective applications. While monocrystalline lithium niobate thin films advance nonlinear nanophotonics, their high inertness limits the design of top-down fabricated nanostructures. Here we present a versatile bottom-up fabrication method based on nanoimprint lithography for achieving polycrystalline lithium niobate nanostructures, and demonstrate its significant potential for nonlinear metasurfaces. The fabrication enables nearly vertical features and aspect ratios of up to 6, which we combine with a novel solution-derived material with high effective second-order nonlinearity of $d_{\text{eff}}$=5 pm/V. On this platform, we demonstrate second-harmonic focusing over a broad spectral range from near-UV to near-IR, increasing the nonlinear signal intensity by up to 34 times. Our method enables the first lithium niobate metalens and expands the field of nonlinear metasurfaces by providing a low-cost, highly scalable fabrication method for engineered nonlinear nanostructures.

\section{Introduction}

Flat optics has opened a new toolbox for controlling light propagation in space. The collective response of carefully engineered sub-wavelength building blocks, called meta-atoms, allows tuning the phase of the incoming wavefront, enabling functionalities such as beam steering, focusing or holography. The combination of metasurfaces with nonlinear materials extends this tunability also to the frequency dimension \cite{Pertsch2020, Fedotova2022}.

This is particularly relevant for the ongoing efforts to reduce the cost, availability and footprint of many spectroscopy tools while extending their detection range \cite{Cai2023, Wang2023, Tseng2020}.
Second-harmonic generation (SHG) is typically the strongest of the nonlinear processes. However, it is only observed in non-centrosymmetric crystalline materials or as surface-SHG, dominant in plasmonic systems. The latter was the first material platform to demonstrate SHG focusing with a geometric-phase metalens made from resonant gold nanostructures \cite{Schlickriede2018}. However, the fabrication robustness and stability at high optical powers are limited due to the roughness sensitivity of surface-SHG and ohmic losses.  As a result, research efforts have focused on investigating alternative, dielectric material platforms. \\
In particular, dielectric materials ideally feature broadband transparency, a high refractive index, and nonlinear crystal properties. 
Non-centrosymmetric metal-oxides are a widely used material platform for nonlinear signal generation. In contrast to e.g. GaAs \cite{Gigli2021}, they feature transparency down to the visible and ultraviolet range. Their high optical power damage threshold allows exploiting the quadratic scaling of SHG intensity with the pump power. However, using these materials to fabricate optical metasurfaces has been challenging due to their chemical and physical inertness during nanostructuring. The first non-centrosymmetric material nanostructured for SHG focusing with a metalens was polycrystalline ZnO, enabling resonance enhanced simultaneous deep-UV generation and focusing for intensity enhancement \cite{Tseng2022}. The commercialization of thin film lithium niobate, LiNbO\textsubscript{3} (LN, $d_{33}$=34 pm/V at $\lambda_{\text{pump}}=1064$ nm \cite{Shoji1997}) led to a significant boost in nonlinear nanophotonics \cite{Boes2023}. While resonant LN metasurfaces \cite{Fedotova2020} and photonic integrated circuits \cite{Zhu2021} have been demonstrated in the last decade, nonlinear wavefront shaping by LN nanostructures is still challenging with single examples utilizing discrete diffraction orders \cite{Carletti2021} or hybrid plasmonic structures \cite{DamgaardCarstensen2021}. The harsh etching processes needed for top-down fabrication of LN require specialized equipment and results in characteristic slanted sidewalls, which prevent achieving high aspect-ratio structures needed for unrestricted wavefront control in particular for the visible wavelength range\cite{Kaufmann2023}. Furthermore, the drastically varying $ \chi^{(2)}$ tensor components in monocrystalline LN conflict with the rotational symmetry requirement in metalens designs for isotropic focusing.\\
However, top-down fabrication processes can be alternatively replaced using the versatile and high-resolution soft nanoimprinting (SNIL) technique  \cite{Kuznetsov2024}, which relies on replicating a silicon master mold design via intermediate polydimethylsiloxane (PDMS) molds. This scalable PDMS replication of designs, with dimensions close to latest state-of-the art electron beam lithography, can be used to directly mold inorganic materials. Compared to capillary stamping \cite{Alarslan2021} and template-assisted dewetting \cite{DasGupta2019}, this leads to superior resolution and high aspect ratios \cite{Modaresialam2021, Verschuuren2017}. To use this technique for nonlinear metasurfaces, it is necessary to develop a solution-derived nonlinear medium. Our recent publications demonstrated imprinted barium titanate metasurfaces \cite{Talts2023, Weigand2024} and motivated the search for materials with higher effective optical nonlinearity. \\
Here, we introduce a novel polycrystalline LN material platform that has been optimized for direct SNIL. This enables highly scalable and flexible fabrication of specifically engineered nonlinear metal-oxide nanostructures. The material properties like refractive index, effective second-order nonlinear coefficient and SHG polarization dependence are investigated. The resulting polycrystalline LN has an effective optical second-order nonlinearity coefficient $d_{\text{eff}}$ of 4.8 pm/V at $\lambda_{\text{pump}}$=880 nm (0.14$\times d_{33}$ of LN). The ease of nanostructuring this material combined with its high second-order nonlinearity enables cost-efficient and highly scalable fabrication of nonlinear metasurfaces. We demonstrate second-order nonlinear geometric-phase metalenses, which work in a broad spectral range from the near-UV to the IR ($\lambda_{\text{pump}}$= 760-1550 nm). The over 30 times enhanced nonlinear intensity of the SHG at the focal point of the metalens can benefit hardware development for applications in nonlinear microscopy and spectroscopy, nonlinear holography for anti-counterfeiting technology or tunable flat optics.

\section{Results}
\subsection{Sol-gel LiNbO\textsubscript{3} for imprint lithography}
Direct soft imprint lithography enables large-scale nanopatterning of any material that can be molded with a plastic elastomer mold. The required high optical quality of the imprinted metasurfaces necessitates a homogeneous densification during the process. The sol-gel synthesis recipe is therefore specifically tuned for low viscosity and high stability in ambient conditions. Metal ethoxide precursors were chosen for the Li- and Nb- ion sources to reduce the introduction of additional non-volatile chemical elements. To increase the chemical stability and imprinting conformity, gelation agents, anhydrous solvents and a lower pH of the solution were employed. \cite{Edmondson2019,Simonsen2009,Danks2016}. The final recipe used for the imprinting has a concentration of 0.125 mol/l for Nb\textsuperscript{+5} ions and 0.145 mol/l for Li\textsuperscript{+} ions  to compensate for the different evaporation rates of the two elements, ensuring correct stoichiometry in the final sample. Crystalline films as well as nanostructures were formed by annealing the samples at 600 \textdegree C for 5 hours. Synthesis details are included in the Methods, while detailed X-ray diffraction (XRD) results are given in the  Supplementary Section 2. \\
\begin{figure}[h!]
    \centering
    \includegraphics[width=\textwidth]{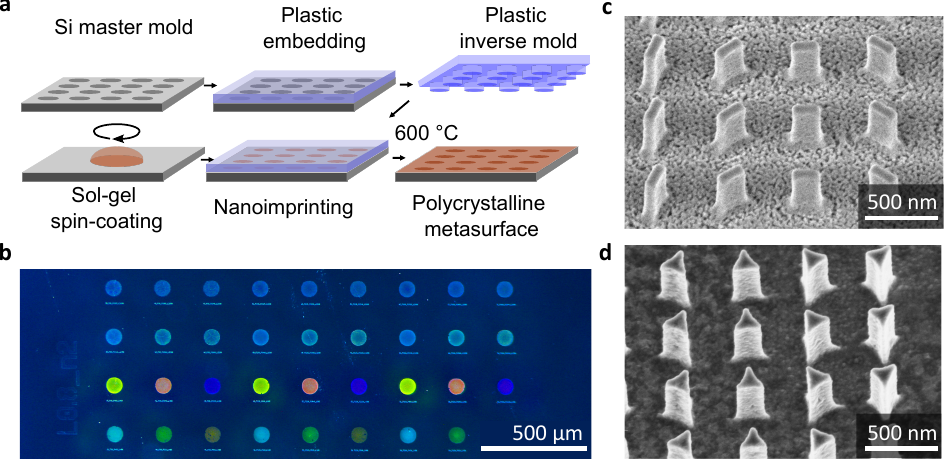}
    \caption{Direct soft nanoimprint lithography (SNIL) with solution derived lithium niobate (LN). a) Schematic of the SNIL process flow adapted to imprinting polycrystalline metal-oxides. b) Dark field optical microscopy image of an array of LN metalenses with varying periodicity (480 nm - 2.24 {\textmu}m), unit cell dimensions of 150 nm - 800 nm, and a fixed height of 440 nm. c) Features of an imprinted C2-symmetric metalens imaged at 30{\textdegree} tilt. d) Features of an imprinted C3-symmetric metalens imaged at 30{\textdegree} tilt. }
    \label{SNIL}
\end{figure}
Using the direct SNIL technique (illustrated in Fig. \ref{SNIL}a, details included in Methods), we fabricated a large set of metalenses, shown in the optical darkfield image in Fig. \ref{SNIL}b. 
The different metalenses have varying periodicities between 480 nm to 2.24 {\textmu}m with minimum structure dimensions of 70 nm, resulting in aspect ratios of up to 6. Here, we demonstrate fin-shaped (Fig. \ref{SNIL}c) and triangle-shaped (Fig. \ref{SNIL}d) nanostructures, which are commonly used features in geometric-phase metalenses. Having comparable results with the state-of-the-art fabrication capabilities, the SNIL approach is suitable for low-cost, wafer-scale fabrication of metasurfaces using commercially available tools for direct imprinting of inorganic structures from sol-gels \cite{Verschuuren2017}. \\
To ensure that the imprinted structures are crystalline LN, we evaluate the smallest imprinted meta-atoms with a triangular prism design (base length of 200 nm, height of 300 nm) in a high-resolution transmission electron microscope (HR-TEM). The triangular unit cell structures shown in Fig. \ref{SNIL}d were cut via focused ion beam (FIB) milling into approximately 40 nm thin lamellae for TEM analysis. The cross-section image taken in scanning TEM mode, Fig. \ref{TEM}a, shows that these structures have an aspect ratio greater than 2 and exhibit low porosity. Energy-dispersive X-ray spectroscopy was used to confirm that the structures are composed of Nb and O atoms (Supplementary Fig. SI 3). The remaining residual film is approximately 50 nm thick and shows significantly higher porosity than the imprinted structures. This might be caused by the volumetric shrinkage occurring mainly in the nanostructures during the sol-gel solidification \cite{Modaresialam2021}.\\
\begin{figure}[h!]
    \centering
    \includegraphics[width=\textwidth]{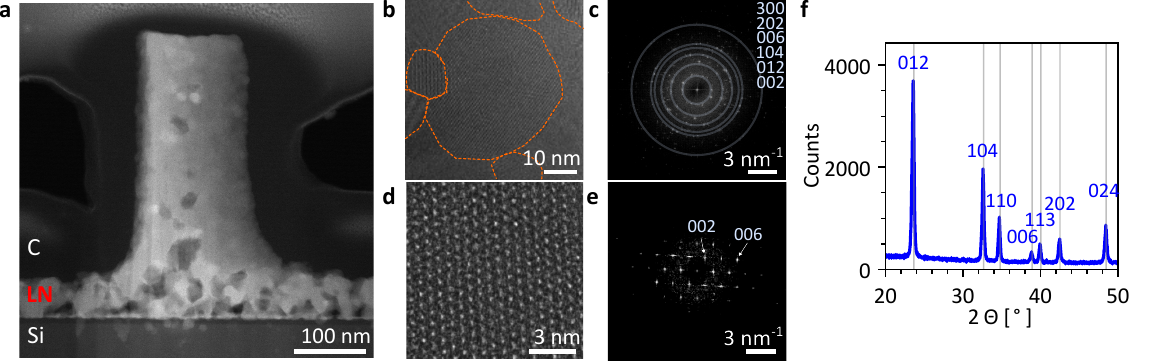}
    \caption{Imprinted LN crystallinity characterization a) Scanning TEM image showing a cross-section of an imprinted nanoprism structure. b) High resolution TEM (HR-TEM) image of the polycrystalline features within the imprinted structure with prominent crystalline domains outlined in orange. c) Fast Fourier transform (FFT) of b) with polycrystalline diffraction patterns matching the lattice spacings of the indicated crystal planes d) HR-TEM image of a single domain within the imprinted structure. e) FFT of d) with diffraction spots matching the lattice of LN. e) Powder x-ray diffraction of films prepared using the same recipe. Reference values plotted in gray lines taken from JCPDS No. 00-020-0631. }
    \label{TEM}
\end{figure}
The HR-TEM allows clear identification of crystalline grains with sizes between 10-30 nm and random orientations (Fig. \ref{TEM}b). The co-presence of different crystalline growth directions within a single nanostructure is further confirmed in the k-space (Fig. \ref{TEM}c), where diffraction patterns of several characteristic crystalline planes of LN can be seen. An analysis of one individual domain (Fig. \ref{TEM}d) with the corresponding k-space image (FFT, Fig. \ref{TEM}e) shows characteristic diffraction points for crystalline planes of LN. Powder XRD measurements of drop-cast polycrystalline films of LN in Fig. \ref{TEM}f additionally verify the presence of crystalline LN in our material.\\
These material characterization results confirm that our solution-based LN is compatible with the direct imprint method for fabricating nanostructures with a non-centrosymmetric crystalline structure. This enables large scale fabrication of nonlinear LN nanostructures with high aspect ratios, low porosity, and nearly vertical, smooth side-walls. These features make this novel polycrystalline metasurface platform competitive with state-of-the-art lithium niobate metasurfaces. \\

\subsection{Second-harmonic generation in polycrystalline LiNbO\textsubscript{3} }
\begin{figure}[h!]
    \centering
    \includegraphics[width=\textwidth]{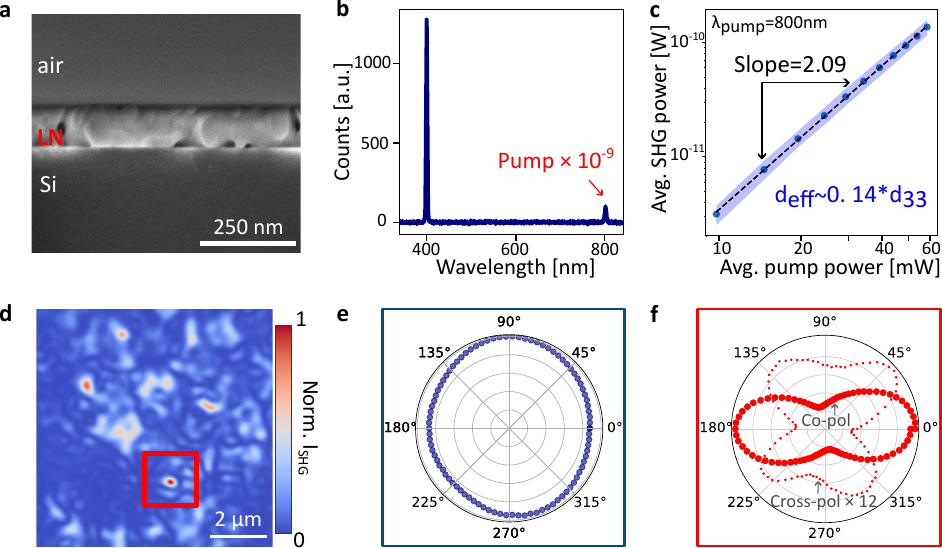}
    \caption{Second-harmonic generation (SHG) in sol-gel LN thin film. a) SEM image of LN film cross-section on Si. b) SHG spectrum measurement ($\lambda_{\text{pump}}=800$ nm, average power 150 mW) with an optical filter attenuating the pump by 9 orders of magnitude. c) SHG power dependence on input fundamental power. Averages of five measurements are plotted in log scale. The effective second-order nonlinearity $d_{\text{eff}}$ is determined from comparison with x-cut monocrystalline LN thin film $d_{33}$ tensor value.  d) Example of a characteristic SHG speckle pattern for $\lambda_{\text{SHG}}=400$ nm at 60 mW pump power. Whole image shown here corresponds to the analysis area for part e. Red box marks an example area used for the analysis of an individual speckle as shown in part f. e) SHG intensity dependence on pump polarization for the speckle pattern shown in d). f) Co- and cross-polarization plots for high intensity SHG speckle corresponding to 3.7 $\mu \text{m}^2$ analysis area as indicated with the red box in d. }
    \label{Film}
\end{figure}

To evaluate the effective second-order optical nonlinearity tensor coefficient ($d_{\text{eff}}$) of the polycrystalline LN, a 120 nm thin film was prepared by spin-coating the solution-based LN in multiple steps. Details of the sample preparation and optical characterization are given in the Methods and Supplementary Section 3, 4 and 5. The sol-gel recipe results in partially porous thin films (cross-section shown in Fig. \ref{Film}a), similar to the residual film of imprinted nanostructures (Fig. \ref{TEM}a), and has a  measured refractive index close to 1.9 in the visible region (Supplementary Fig. SI 2). \\
The SHG from the solution-derived thin film was studied by illuminating the material with a pulsed laser beam in the near-IR. The transmission spectrum of the film being pumped at a wavelength of 800 nm shows the doubling of the fundamental laser frequency to 400 nm and the absence of fluorescence (Fig. \ref{Film}b, pump attenuated by optical filters). Furthermore, the second-order nonlinearity of the signal was confirmed by measuring the intensity of the SHG as a function of the fundamental laser power, verifying the expected quadratic dependence (Fig. \ref{Film}c, logarithmic scale). The effective second-order optical coefficient $d_{\text{eff}}$ at $\lambda_{\text{pump}}$=880 nm was evaluated by comparing the SHG signal from the sol-gel film with the signal from monocrystalline x-cut LN thin film in a commercial 2-photon microscopy setup (details in Methods and the Supplementary Section 6, Fig. SI 5b,c and \cite{SnchezDena2017}). As expected from a randomly oriented nonlinear medium, the effective optical nonlinearity of the solution-derived film is reduced, reaching 14 \% of the monocrystalline $d_{33}$ tensor value strength. This value is close to the maximally achievable 20 \% for randomly oriented domains with bulk LN $\chi^{2}$ tensor values \cite{SnchezDena2017} and thus coherent with the HR-TEM and XRD results showing no dominating orientation (Fig. \ref{TEM}). \\
To evaluate the in-plane domain orientations of the material, a SHG intensity map of the solution-derived LN was recorded in a transmission configuration, featuring the characteristic speckle pattern of polycrystalline materials (Fig. \ref{Film}d). The  SHG at different pump polarizations, generated by a fundamental beam with a peak intensity of 1.4 GW/cm\textsuperscript{2}, is plotted in Fig. \ref{Film}e for the whole area analyzed in Fig. \ref{Film}d. The polarization-dependent plot confirms an averaged isotropic SHG intensity dependence of the thin film. Further polarization decomposition into co- and cross-polarized SHG compared to the pump beam polarization in Fig. \ref{Film}f shows that the high-intensity speckles (red square marking in Fig. \ref{Film}d) have a distinct anisotropic shape similar to x-cut LN thin films (Supplementary Fig. SI 5a).
An analysis of individual imprinted nanostructures, as investigated in the TEM section, revealed a dominant  nonlinear tensor orientation within the single prisms, but no globally preferred orientation in equivalent polycrystalline structures (see Supplementary Section 7).

\subsection{Polycrystalline Lithium Niobate Metalenses}
\begin{figure}[h!]
    \centering
    \includegraphics[width=\textwidth]{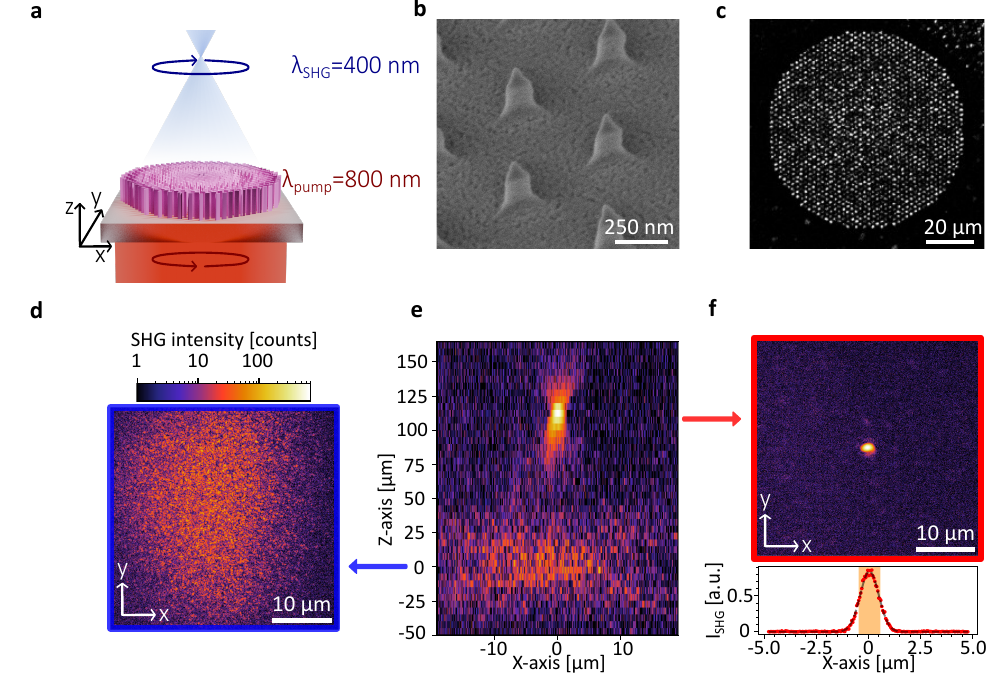}
    \caption{Nanoimprinted nonlinear metalenses for SHG focusing. a) Illustration of the metalens probed in transmission configuration with nearly plane wave, circular polarized illumination. b) SEM image of several unit cells in the hexagonal C3-symmetric metalens at 30° tilt. c) SHG image acquired via raster-scan-based two-photon microscopy of a metalens with larger unit cells. d) SHG at metalens surface (z=0) in x-y plane. e) SHG in x-z- plane (y=0) along the optical axis. f) SHG in x-y plane at the metalens focal point at z=115 {\textmu}m from sample surface. Below is a magnified plot of the projected point spread function of the intensity at the focal point with a  Gaussian fit and a FWHM of 1050 nm, highlighted in yellow. Measurements were done with right-handed circular polarization of pump beam with an average intensity of 0.4 $\text{kW}/\text{cm}^2$ at the sample. d)-f) share the same colorscale.  }
    \label{metalens}
\end{figure}

Geometric phase metasurfaces are used to control the phase and polarization of circularly polarized light. This design is commonly used to create metalenses for robust and broadband focusing, without having to rely on delicate resonance-induced phase shifts. Macroscopic selection rules of the nanostructures and the lattice govern the circularly polarized wave propagation in polycrystalline and amorphous metasurfaces, allowing second-harmonic wavefront control with one-fold and three-fold rotational  symmetry (C3-symmetry) of the meta-atoms \cite{Bhagavantam1972, Li2017}.  These symmetry-based selection rules can be harnessed for devices featuring stacked metasurfaces with different functionalities \cite{Li2017}. Our study focuses on nonlinear metalenses with C3-symmetry to minimally influence the linear wavefront propagation. The beam shaping is induced by the geometric phase, which relates the geometric rotation angle of the unit cell to the phase delay acquired by circularly polarized light with opposite polarization handedness to the pump (illustrated in Fig. \ref{metalens}a): 

\begin{equation}
\theta (r) = \frac{2\pi}{(m+1)\lambda}(\sqrt{f^2 + r^2}-f)
\label{PBlens}
\end{equation}
Where $\theta$ is the geometric unit cell rotation, $m$ is the order of the higher harmonic process in the metalens ($m=2$ for SHG), $\lambda$ is the steered wavelength, $f$ is the focal distance and $r$ is the in-plane distance from the metalens center. \\
The unit cells of the metalens are shown in Fig. \ref{metalens}b, consisting of triangular prisms with a base length of 150 nm, a periodicity of 480 nm and a height of 300 nm, arranged in a hexagonal lattice with rotations following equation \eqref{PBlens}. The SHG from an imprinted metalens with 80 {\textmu}m diameter was confirmed with 2-photon microscopy, where the prism nanostructures clearly stand out from the residual thin film (Fig. \ref{metalens}c). We ensured that the metalens performance is not affected by resonances by measuring wavelength-dependent SHG of the metalens and comparing it to that of an unstructured, 120 nm thin film (Supplementary Fig. SI 7). We characterized the second-harmonic wavefront generated by the imprinted nonlinear metasurface by recording the focal plane of a collection objective with a camera as it moved along the z-axis away from the metalens (setup details in Supplementary Fig. SI 4). A lens with a focal length of 100 mm was used to loosely focus the circularly polarized pump onto the metalens surface, generating second-harmonic signal from an area of 38 {\textmu}m diameter (D), where D was determined as the 1/e$^2$ SHG intensity drop at the surface of the metalens (z=0) from Fig. \ref{metalens}d. The evolution of the SHG wavefront is shown along the xz-plane (y=0) in Fig. \ref{metalens}e, with the surface of the metalens at z=0 {\textmu}m and the focus of the metalens being clearly visible at z=115 {\textmu}m. The surface of the metalens, xy-plane at  z=0 {\textmu}m depicted in Fig. \ref{metalens}d, shows the characteristic SHG speckle pattern similar to Fig. \ref{Film}d. This speckle pattern converges into a homogeneous focal spot  at 115 {\textmu}m from the surface, without any higher diffraction orders visible in the xy-plane (Fig. \ref{metalens}f). This results in a numerical aperture (NA) of 0.16 for the metalens under investigation. While we confirm the geometric phase induced handedness-change of the focused SHG polarization compared to the pump using an analyzer, the SHG at the focal point shown here was clearly distinguishable without polarization-selective optical components in the collection path.
At a pump excitation wavelength of 800 nm, the SHG focal spot has a measured full-width at half-maximum (FWHM) of 1050 nm. The small difference from the theoretical diffraction limit of 907 nm (FWHM) for Gaussian beams ($\frac{4\lambda_{\text{SHG}}f}{\pi D}\times\frac{\sqrt{-2\ln{0.5}}}{2}$) can be attributed to fabrication-induced deviations from the analytical phase profile \cite{Tseng2022}. Importantly, the experimental resolution of the nonlinear metalens nevertheless exceeds the theoretical limit of the fundamental beam (1814 nm FWHM), leading to superior resolution compared to linear lenses.

When metalenses with C2-symmetric unit cells are measured in the same setup, no SHG focusing is detected (details available in Supplementary Fig. SI 8). This confirms that neither the residual LN layer nor the optical components in the beam path contribute to the observed SHG focusing.

\begin{figure}[h!]
    \centering
    \includegraphics[width=\textwidth]{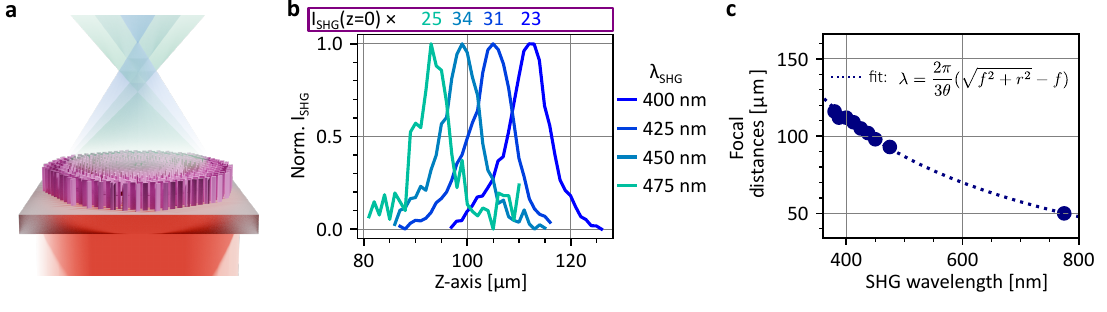}
    \caption{Broadband SHG focusing. a) Illustration of the metalens' chromatic aberration. b) Normalized SHG intensity along the z-axis, displaying the focal point for different SHG wavelengths between 400 and 475 nm with corresponding power density enhancement indicated above. c) Measured broadband SHG focal distance, fitted with chromatic aberration as expected for nonlinear geometric-phase metalenses. Fit extrapolated from experimental data points.}
    \label{fig6}
\end{figure}

The geometric-phase metalens design features intrinsic chromatic aberration, as illustrated in Fig. \ref{fig6}a.  Fig. \ref{fig6}b shows the SHG intensity at the metalens focal points for different wavelengths. Notably, since the metalens does not rely on resonances, all wavelengths feature similar intensity enhancement factors of up to 34 compared with the inhomogeneous, speckle-like SHG at the metalens surface, as indicated on top of Fig. \ref{fig6}b (details included in Supplementary Fig. SI 9 and Table SI 2).  The slightly varying intensity enhancements of the SHG at the focus are likely caused by the difference of wavefront-shaping efficiency by the unit-cell height at different wavelengths. Comparing our results with previous work on polycrystalline ZnO geometric-phase metalenses \cite{Tseng2022}, the intensity enhancement here is nearly a factor of 5 higher (accounting for the difference in SHG polarization detection). Remarkably, our design relies neither on Mie-resonances nor on optimized phase profiles for a selected wavelength. The outstanding intrinsic material nonlinearity and the broadband functionality of the lens enable to experimentally confirm the wavelength-dependence of the focal distance. The results closely follow the analytical design formula for the second-order nonlinear phase propagation introduced in equation \eqref{PBlens}, as shown in Fig. \ref{fig6}c. By measuring the metalens properties in the near-IR with $\lambda_{\text{pump}}$=1550 nm, we further confirm its broadband functionality, spanning SHG focusing from the near-UV below 400 nm to the near-IR above 750 nm. Numerous studies suggest that the performance can be further improved by optimizing the metalens design for higher focusing efficiency, or by leveraging lattice resonances if even stronger SHG is needed at fixed wavelengths \cite{Gigli2021,Tseng2022,Reineke2022}.  \\

\section{Discussion}
In this work, we present a novel solution-derived LN optimized for achieving nonlinear optical metasurfaces via a highly scalable and cost-effective imprint fabrication process. The effective second-order nonlinearity coefficient $d_{\text{eff}}$ of the polycrystalline LN sol-gel is 4.8 pm/V (14 \% of $d_{33}$), which is close to the theoretical maximum of 6.6 pm/V. Combining this material with direct nanoimprint lithography paves the way for high optical quality nonlinear nanostructures, which we use to demonstrate the outstanding frequency conversion efficiency of the solution-derived LN across its broad transparency range. Our work showcases nearly vertical side-walls and aspect ratios of up to 6, which are essential for investigating previously fabrication-limited metasurface designs in non-centrosymmetric metal-oxides. This method is compatible with a wide range of substrate materials and topologies, and the synthesis route enables to further enhance the nonlinear properties through doping \cite{Chen2021, Lukowiak2014}, the integration of quantum dots \cite{Maikhuri2015} or plasmonic nanoparticles \cite{Ali2018}. The interdisciplinary combination of nonlinear optical materials' engineering with direct nanoimprint lithography has the potential to redefine the capabilities and applications of nonlinear metasurfaces.\\
We demonstrate this by fabricating the first geometric-phase LN metalens with C3-symmetry, enabling SHG focusing and intensity enhancement at the focal spot across a broad spectral range from the near-UV to the near-IR. This approach allows for the exploration of more complex metalens designs with enhanced light-matter interaction, including Mie resonances, collective lattice coupled resonances, Fano resonances and bound states in the continuum \cite{Pertsch2020}. Beyond SHG, these polycrystalline metasurfaces also provide a novel, highly scalable platform for studying and tuning the inverse process of SHG, spontaneous parametric down-conversion, for quantum light sources or for utilizing the Pockels effect for GHz-speed free-space modulators.   

\section{Methods}
\subsection{Sol-gel Synthesis}
Lithium niobate sol-gel was synthesised by adapting a previously published recipe \cite{Takahashi2004}. A gelation agent (acetylacetone) and anhydrous solvents were used to increase the chemical stability. Additionally, we observed that lowering the pH of the sol-gel solution was necessary to achieve high conformity during molding and avoid sedimentation of impurity phases. The protonated OH\textsuperscript{-} groups of the metal-alkoxide precursors are less prone to condensation and precipitation before nanostructuring due to repulsive electro-static forces \cite{Simonsen2009,Danks2016}. The exact chemicals used in the synthesis process are provided in Supplementary Table SI 1. 1 mol/l lithium ethoxide solution in ethanol (Sigma-Aldrich) was diluted to 0.28 mol/l in anhydrous n-propnaol. High purity niobium ethoxide (Sigma-Aldrich) was stabilized in acetylacetone with 1:3 Nb\textsuperscript{+5}:acetylacetone ratio. The lithium ethoxide and niobium ethoxide solutions were mixed to result in 0.20 mol/l Nb solution with 1:1.2 molar ratio of Nb\textsuperscript{+5} to Li\textsuperscript{+1} due to higher volatility of Li\textsuperscript{+1}. To improve the stability and shelf-life of the resulting mixture was diluted with glacial acetic acid to 0.125 mol/l Nb concentration. Polycrystalline thin films were prepared on plasma treated silicon (100 orientation, p-type doped) and fused quartz substrates by spin-coating 9 layers of the resulting sol-gel mixture (Step 1: 500 rpm/s, 500 rpm, 5 sec; Step 2: 1000 rpm/s, 1000 rpm, 40 sec) with 30 minute 300 \textdegree C intermediate annealing after 3 deposition layers and final annealing at 600 \textdegree C for 5 hours ( 5 \textdegree C/min ramp up and down). Samples for powder XRD characterization were prepared by drop-casting 30 \textmu l of sol-gel solution onto a Si substrate and annealing at the same conditions. 

\subsection{Direct Soft Nanoimprint Lithography}
Hybrid PDMS molds were prepared as described in previous publications \cite{Talts2023, Weigand2024}. In short, standard electron-beam lithograhpy with hydrosiloxane resist was used to prepare master molds with the metasurfaces without any etching process.
To avoid bonding between the master mold and intermediate PDMS mold, the surface of the silicon mold was treated with a fluorination agent vapor ((heptadecafluoro-1,1,2,2-tetrahydrodecyl)trichlorosilane, Gelest Inc) directly after oxygen plasma treatment. Hybrid PDMS molds were prepared by spin coating (1000 rpm/s, 1000 rpm, 40 sec) an approximately 100 \textmu m film of h-PDMS (Gelest Inc) onto the Si substrate. The h-PDMS covered master molds were preheated at 70 \textdegree C for 10 min before being embedded in an approximately 5 mm thick layer of soft PDMS (Sylgard 184, Sigma-Aldrich). Upon degassing in a vacuum chamber with 250 mbar, the hybrid PDMS molds were cured in a 65 \textdegree C oven for 12 h.  Resulting inverse molds were degassed in 250 mbar vacuum for 10 minutes before being used for direct imprint with LN sol-gel solution.  The metasurfaces were fabricated using SNIL on plasma treated p-type silicon and fused quartz substrates for characterization and optical measurements, respectively. 
The solvent was left to evaporate through the PDMS mold for 3 hours on a 60 \textdegree C hot plate before detaching from the sample. Imprints were annealed following the same protocol as the  thin films (5 \textdegree C/min ramp, 600 \textdegree C, 5 h).  The observed linear shrinkage factor from master mold structures to imprinted and annealed polycrystalline structures was approximately 50$\%$.

\subsection{Material Characterization}
Crystallinity of the resulting films were characterized with powder XRD (PANalytical X’Pert PRO MRD Ge monochromator for $K\alpha_{1}$) and HR-TEM (JEOL JEM F200). Formation of 1:1 stoichiometry LN was confirmed and impurity phases detected by comparing the diffraction peaks of out-of-plane oriented crystallographic planes from drop-casted films with reference values (Supplementary Material Fig. SI 1 ).
For the SHG analysis of LN thin-films, the sol-gel was deposited on Si substrates for cross-section imaging and two-photon microscopy, or on fused silica for transmission measurements.

\subsection{Nonlinear optical measurements}
2-photon microscopy of the imprinted metalens was done with a commercial set-up (Leica TCS SP8 Multiphoton microscope) with 20X objective (HC PL APO CS2, 20x/0.75 DRY Zeiss) in reflection configuration with $\lambda_{\text{pump}}$=880 nm and SHG filter range set to 435-445 nm.\\
Quantitative second-harmonic generation characterization of the thin films, imprinted individual nanostructures and imprinted metalenses was done using a custom-built automatized free-space optical setup as in our previous work \cite{Timpu2019, Saerens2019, Talts2023}. Characterization in the near-IR pump spectral range was done with a Ti:Sapphire laser and SHG was separated with 2 colored glass bandbass filters (Thorlabs, FGB39). To ensure the detected SHG was from the sample, an additional long pass filter (Thorlabs, FELH0750) is placed after the $\lambda/4$ waveplate, before the excitation lens, following the beam path. Metalens focusing characterization in the IR was done with a fiber laser (Menhir Photonics, Menhir-1550) with SHG selected with 2 short-pass filters (Thorlabs, FESH0900). The sample was illuminated with a 50 mm focal distance lens for characterization of the thin films and individual nanostructures (beam diameter of 24 {\textmu}m). A 100 mm focal distance lens was used to loosely focus the pump beam for characterization of the nonlinear metalenses in the near-IR range (beam diameter of 38 {\textmu}m). To characterize the metalenses in the IR range, a lens with f = 50 mm, defocused  by 1 mm, was used due to the detection camera being limited to sub-1100 nm wavelength range and lower pump peak power output.  Optical setup details are included in Supplementary Fig. SI. 4.
\backmatter

\section*{Supplementary Information}
Supplementary information includes Table SI 1 and Table SI 2, Figures SI 1 - 9. 

\section*{Acknowledgments}
The authors thank the Scientific Centre for Optical and Electron Microscopy (ScopeM),
the Binning and Rohrer Nanotechnology Center (BRNC), the FIRST cleanroom and DMATL
X-ray service platfrom at ETH Zurich
for technical assistance. This work was supported by the Swiss National Science Foundation
SNSF (Consolidator Grant 213713 and Grant 179099) as well as the European
Union’s Horizon 2020 research and innovation program from the European Research
Council under the Grant Agreement No. 714837 (Chi2-nanooxides). H.C.W. acknowledges
financial support from the Physics Department at ETH Zurich.
\bibliography{references}



\section*{Supplementary material}

# Scalable Lithium Niobate Nanoimprinting for Nonlinear Metalenses

Ülle-Linda Talts[1†], Helena Weigand[1†], Irene Occhiodori[1], Rachel Grange[1]*

[1]Optical Nanomaterial Group, Institute for Quantum Electronics, Department of Physics, ETH Zurich, Zurich, CH-8093, Switzerland.

[†]These authors contributed equally to this work.

# Supplementary Information

## Contents

## 1. Details of chemicals for polycrystalline lithium niobate

Table SI 1 includes the exact details of chemicals used for polycrystalline lithium niobate ($LiNbO_3$, LN) synthesis.

*Table SI 1. Details of the chemicals used for solution-derived lithium niobate synthesis.*

| Chemical | Product code | Supplier |
|---|---|---|
| Nb(V) ethoxide | AA1468909 | Fisher Scientific AG |
| Acetylacetone | P7754 | Sigma Aldrich Chemie GmbH |
| 1M lithium ethoxide in ethanol | 400246 | Life Technologies Europe B.V. |
| Glacial acetic acid | A6283 | Sigma Aldrich Chemie GmbH |
| Anhydrous 2-propanol | 5.89584 | Sigma Aldrich Chemie GmbH |

## 2. X-ray diffraction

Powder x-ray diffraction (XRD) measurements were used to evaluate the quality of the sol-gel precursor used for preparing the polycrystalline LN film. A reference sample was prepared by drop-casting 30 μl of sol-gel solution on a crystalline Si substrate and exposing the reference sample to the same annealing conditions as the thin films and imprinted metasurfaces (5h at 600 °C, ramp

rate 5 °C/min). The diffraction peaks were compared to reference values for polycrystalline LiNbO$_3$, LiNb$_3$O$_8$ and Nb$_2$O$_5$ (Fig. SI 1.a). A shelf-life of 1 week at room temperature in a nitrogen purged storage cupboard was determined by noticeable appearance of a LiNb$_3$O$_8$ impurity phase (Fig. SI 1.b).

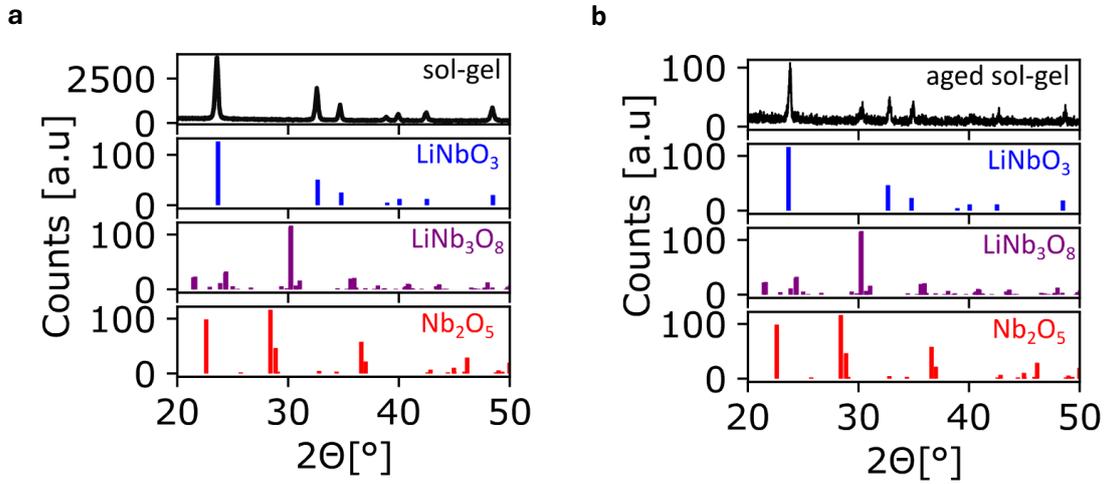

*Figure SI 1. Comparison of powder XRD results for polycrystalline LiNbO$_3$ prepared at different times since synthesis of the sol-gel. a) Sol-gel prepared the day of sample preparation b) Sol-gel prepared a week before sample preparation. Results are compared to the following references: LiNbO$_3$ - JCPDS card no 20-0631 ; LiNb$_3$O$_8$ - JCPDS card no 36-0307 ; Nb$_2$O$_5$ - JCPDS card no 30-0873.*

## 3. Refractive index of thin film

The polycrystalline thin film was prepared on a Si substrate following the same protocol for cross-section imaging (Fig.3a) and for measuring the effective refractive index with ellipsometry (Fig. SI 2). Using the Maxwell Garnett approximation and estimating the medium to consist of only crystalline LN and air inclusions, an approximate porosity of 30 % can be inferred. Due to the air inclusions and the lower atomic density within the polycrystalline material grain boundaries [1], the film has a lower refractive index compared to bulk LN.

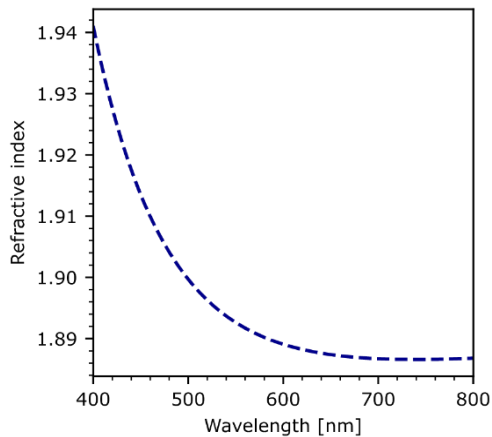

*Figure SI 2. Measured refractive index of solution derived polycrystalline LN film of thickness 120 nm.*

## 4. Elemental analysis in nanostructures

Energy dispersive X-ray (EDX) analysis was used to confirm the element composition of the imprinted triangular nanoprisms. The measurements were done on a 40 nm thin lamella prepared for TEM measurements. The imaged area is shown in Fig. SI 3a, distribution of Nb and O atoms is shown in Fig. SI 3b and 3c, respectively. Li distribution cannot be detected using this method.

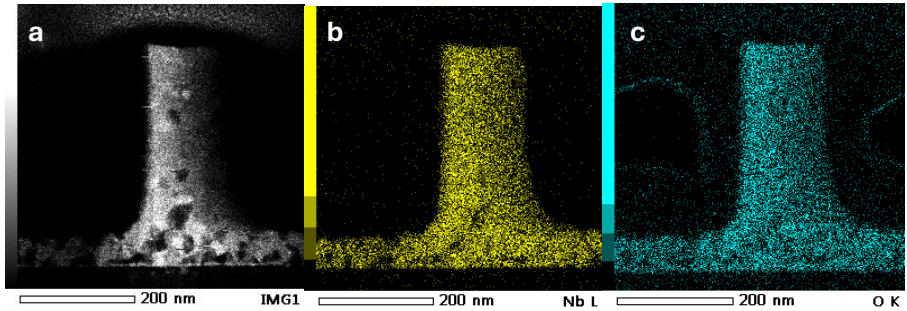

*Figure SI 3. EDX results from imprinted polycrystalline LN -nanostructures . The imaged area a) -without element selection, b) with showing only Nb Kα1 emission at 2.166 keV , c) with showing only O Kα emission at 0.523 keV.*

## 5. Custom optical setup details

Fig. SI 4 details the second-harmonic generation (SHG) measurement setup for characterizing the pump power dependence and emission spectrum (Fig. SI 4a), in-plane co- and cross-polarization features (Fig. SI 4b), nonlinear wavefront shaping by excitations with a right-handed circularly polarized light with pump laser in the near-IR range (Fig. SI 4c) and in the IR (Fig. SI 4d).

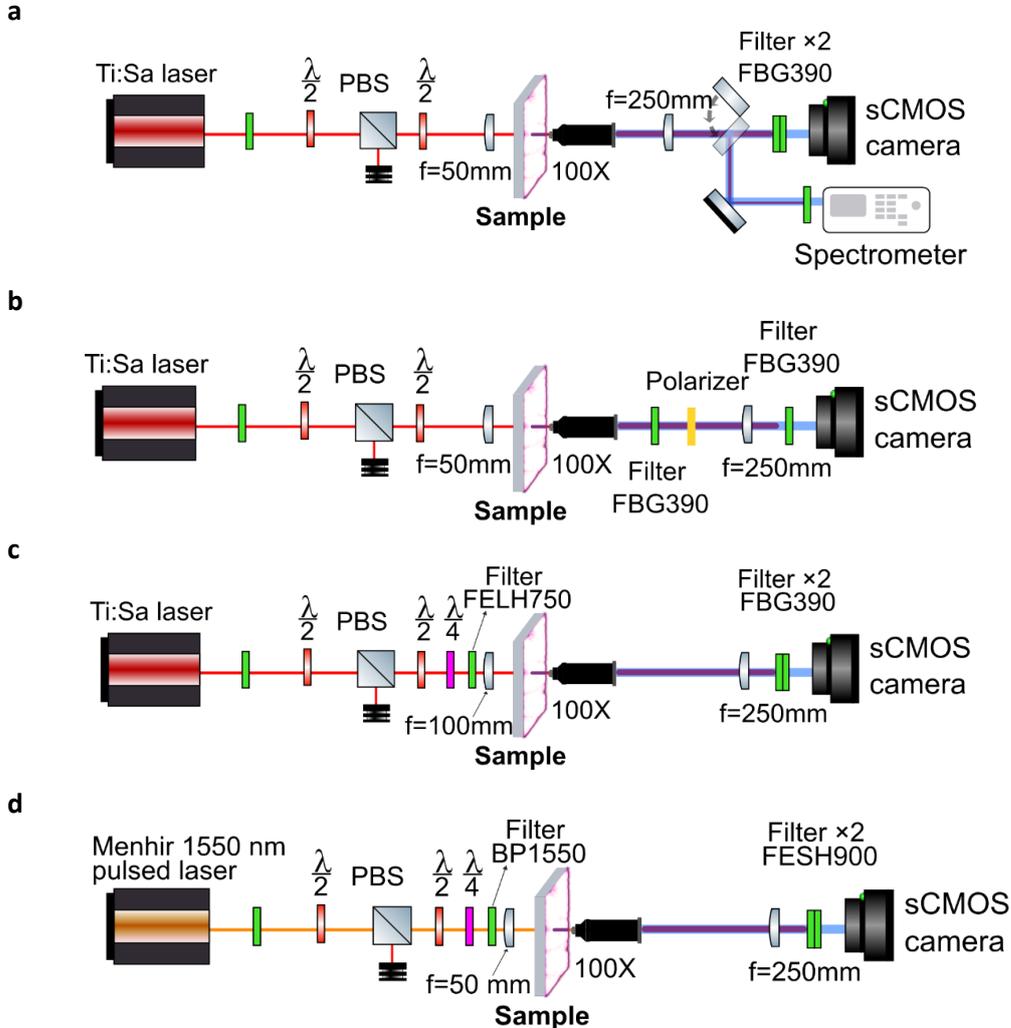

*Figure SI 4: Schematics of the optical setups used for characterizing the SHG from polycrystalline lithium niobate. a) Setup for SHG characterization of thin film and individual nanoprism. b) Setup for analysis of domain orientation in thin film and individual nanoprisms via SHG polarization characterization. c) Setup for metalens SHG characterization in the $\lambda_{pump}$=760-950 nm range. d) Setup for metalens SHG characterization with $\lambda_{pump}$=1550 nm, where a 50 mm lens before the sample was defocused 1 mm towards the camera for increasing the beam width at the metalens surface.*

## 6. SHG from monocrystalline x-cut LiNbO$_3$

The second-order behaviour of the sol-gel LN can be analyzed with respect to the behaviour of a monocrystalline thin-film of LN. For example, the polarization dependence of sol-gel nanostructures can be better understood if compared to the same measurement of monocrystalline LN on quartz. To this end, we measured a monocrystalline x-cut lithium niobate thin film on quartz (NanoLN) as a reference in the same characterization setup. Co-and cross-polarized SHG dependence on the pump laser was measured (Fig. SI 5 a), which shows the comparative strength of the monocrystalline $d_{33}$ tensor component.

4

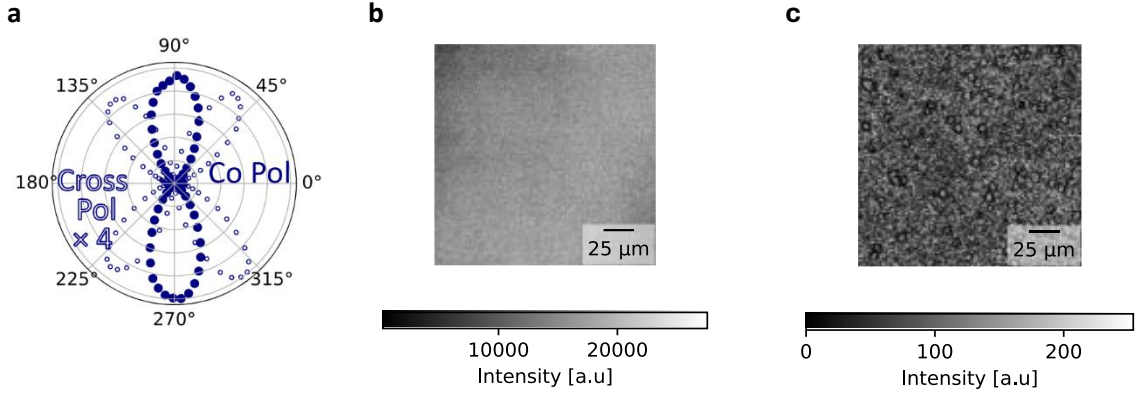

*Figure SI 5 Characterization of monocrystalline x-cut lithium niobate thin film SHG. a) Co- and cross-polarization plots for SHG intensity dependence on the pump polarization with λ$_{pump}$ = 800 nm. Comparative measurement of SHG in reflection configuration: b) 600 nm x-cut lithium niobate thin film with pump aligned to the d33 tensor, c) 120 nm solution-derived polycrystalline lithium niobate thin film.*

Furthermore, a 600 nm thick monocrystalline LN film on a 2 μm SiO$_2$ buffer layer on top of a Si substrate was measured in reflection in a two-photon-microscope (Leica TCS SP8 Multiphoton microscope) over a large area of 180x180 μm². To determine the effective second-order nonlinearity tensor d$_{eff}$ of the polycrystalline thin film, we analyzed the same area on a sol-gel LN film on Si, which had a thickness of 120 nm. Both SHG images are plotted in Fig. SI 5 b and c, respectively. Notably, the polarization of the pump beam has been aligned to the strong monocrystalline $d_{33}$ tensor component. While the monocrystalline SHG is homogeneous, the sol-gel SHG exhibits the typical speckle pattern expected for polycrystalline media. By taking the average over the whole area, we make sure to analyze the effective nonlinear strength of the randomly distributed domains, similarly as for the metalens illumination. The relative strength of the effective nonlinearity in sol-gel can be estimated by comparing the strength of both samples via an equation derived for the Makers-Fringe method [2]. It can be simplified to an experiment-specific factor A, sample specific parameters such as the thickness *L*, the effectively probed second-order tensor χ$^{(2)}$, and the refractive index n at the second-harmonic wavelength:

$$P_{SHG} = A_{exp} \frac{L^2}{n^2} \chi^2_{eff} P^2_{pump}$$

Considering that the experimental conditions and the incident power are the same, we can calculate the following ratio:

$$\frac{\chi^{(2)}_{\text{sol-gel,Si}}}{\chi^{(2)}_{\text{mono.,Si}}} \approx \sqrt{\frac{P_{SHG,sg}}{P_{SHG,mono}}} \cdot \frac{L_{mono} \cdot n_{sg}(2\omega)}{L_{sg} \cdot n_{mono}(2\omega)} \approx 14\%$$

While the $\chi^{(2)}_{\text{mono.,Si}}$ here probes the $d_{33}$ tensor component (34 pm/V), the $\chi^{(2)}_{\text{sol-gel,Si}}$ probes all tensor components simultaneously due to the random orientation of the grains. This gives a theoretically maximally achievable strength of 19.5% of the $d_{33}$ tensor component. Therefore, the 14% measured experimentally here hints to the impressively high SHG efficiency achieved with our solution-based LN.

## 7. SHG from individual nanoprisms

To determine if the nonlinear properties of the imprinted nanostructures are consistent with those of the layer-by-layer deposited thin-film, we investigate the SHG characteristics of three individual imprinted triangular prisms (Figure SI 6a). The results presented here focus on individual prisms with a base side length of 360 nm and a height of 300 nm, although similar characteristics were observed for structures with larger base dimensions. The remaining residual film, visible in Fig. 2a, exhibits a SHG signal 2 orders of magnitude lower than that of the nanoprisms, making the structures clearly visible (Figure SI 6b). The individual nanostructures show a clear dependence of SHG intensity on pump polarization, with an extinction ratio of 96% relative to the maximum intensity (Figure SI 6c). However, on a global scale, the nanoprisms show no preferential crystal orientation neither along the geometry of the structure nor the substrate, as evidenced by the polarization dependence of SHG for three different prisms shown in Figure SI 6c. Additionally, the strength of the SHG varies significantly, indicating that different nonlinear tensor components are excited depending on the random distribution of domain orientations within the nanoprism. To maximize the SHG from each metaatom within a metasurface, circularly polarized pump beams are chosen to ensure all randomly oriented domains of this polycrystalline material are excited.

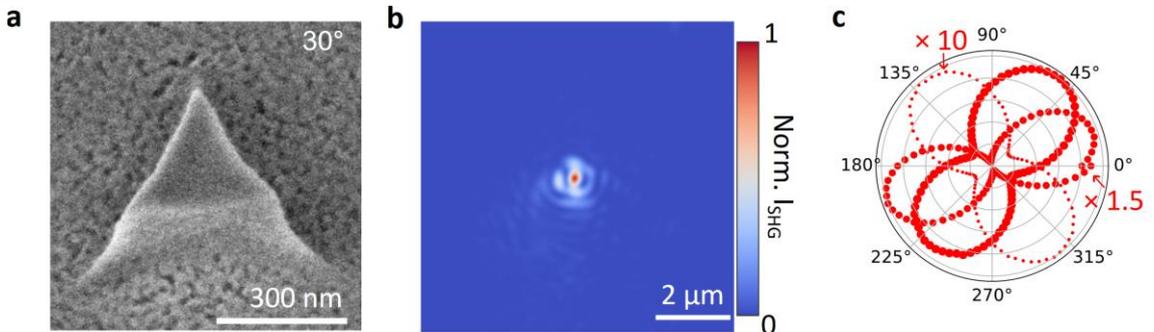

*Figure SI 6 SHG from individual polycrystalline nanostructures. a) SEM of triangular prism unit cell imaged under 30°angle . b) Example of a characteristic, point source like SHG feature for λSHG = 400 nm at 200 mW pump power. The image shown here corresponds to the analysis area for part c. c) SHG intensity dependence on pump polarization for three individual nanoprisms.*

## 8. Wavelength dependent SHG

SHG intensity dependance on pump and upconverted light wavelength was measured by measuring the metalens and the 120 nm polycrystalline thin film in the same optical configuration (set-up details in Fig. SI 4c). The results in Figure SI 8 show how the metalens has a significantly lower SHG compared to the film due to the presence of less nonlinear material in the metasurface. Slight increase in the relative SHG enhancement at lower wavelengths was consistent in lenses with different periodicities. This is hypothesized to stem from higher scattering in the thin film at smaller SHG wavelengths due to the porosity.

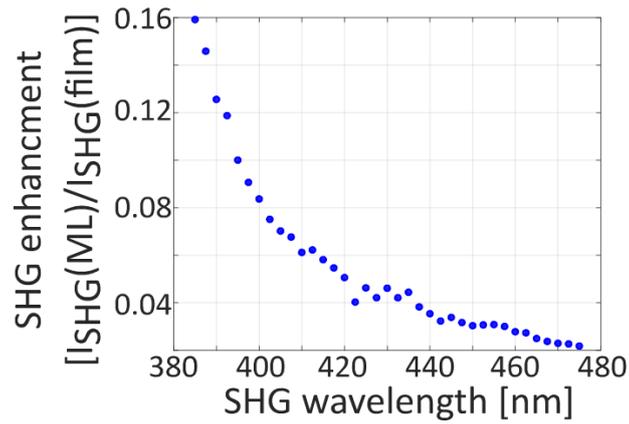

*Figure SI 7 Wavelength dependent SHG relative comparison between analyzed metalens and 120 nm thin film with constant 100 mW average pump power.*

## 9. Wavefront shaping for linear and SHG

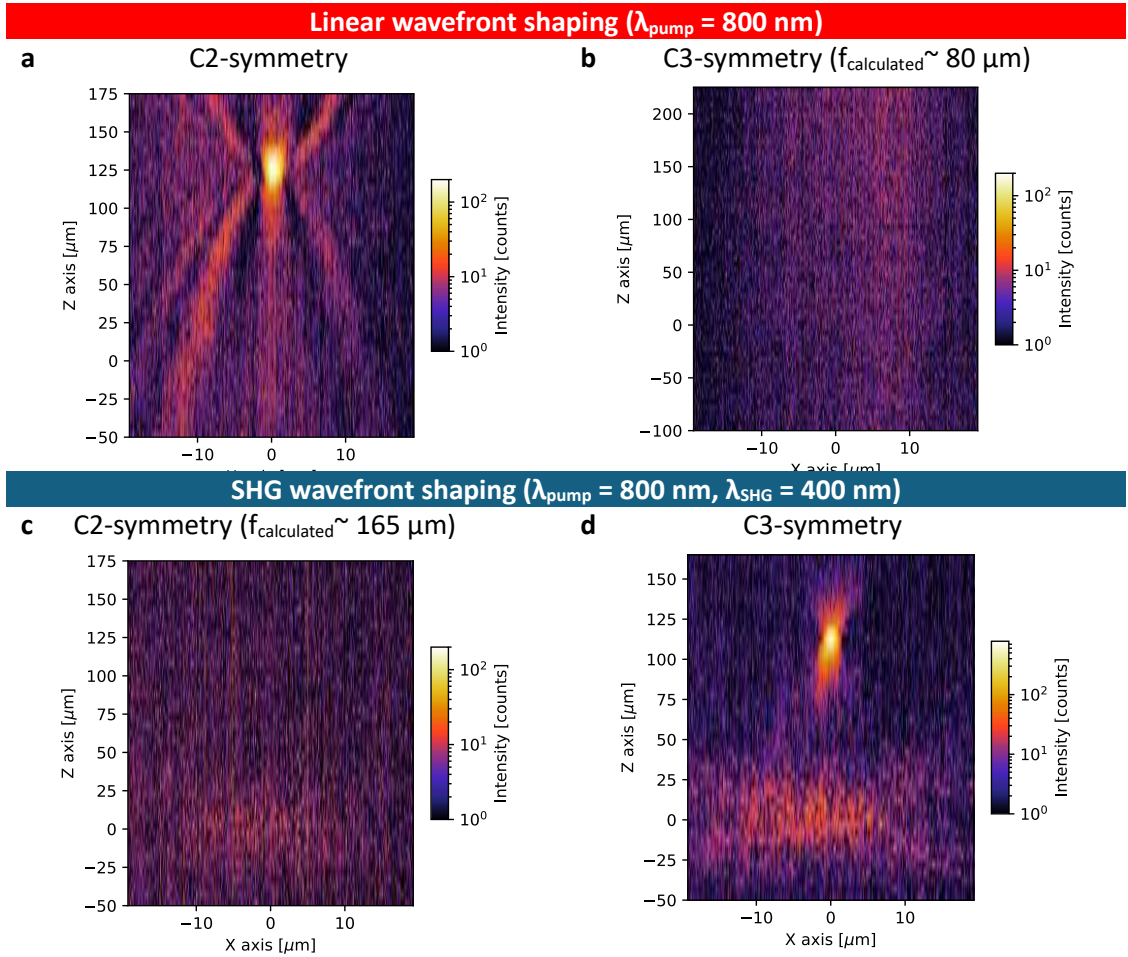

*Figure SI 8 Control measurements of geometric-phase metalens focusing in linear and SHG regime. a) Fundamental beam focusing by C2-symmtery (rectangular) unit cells with cuboids as meta-atoms. b) Fundamental beam propagation by C3-symmetry (triangular) unit cells with prisms as meta-atoms. c) SHG at the C2-symmetry metalens surface with no focusing. d) SHG at the C3-symmetry surface with focusing.*

To ensure the focusing was due to the nonlinear metalens and not from the residual film nor any of the optical components in the path, control measurements on C2- and C3-symmetry unit cell design metalenses were performed (Fig. SI 7). For the linear part, the C2-symmetric lens shows a pronounced focusing, while the C3-symmetric lens does not alter the propagating wavefront, as expected from theory [3]. For the second-harmonic wavefront shaping, the C2-symmetric lens does emit SHG radiation from the LN material, but does not focus it, which is just the case for the C3-symmetric metalens. Note that the focal distance of the symmetry-systems was designed differently:

All C2-symmetry metalenses are designed to focus light of 0.8 µm wavelength in a SHG process (m=2) at 80 µm from the surface, resulting in focusing at 125 µm for linear focusing (m=1) at 0.8 µm.

$$\Theta = \frac{2\pi}{(m+1)\lambda}\left(\sqrt{f^2 + r^2} - f\right)$$

$$\Theta = \frac{2\pi}{3\lambda_{fund}}\left(\sqrt{80^2 + r^2} - 80\right) = \frac{2\pi}{2\lambda_{fund}}\left(\sqrt{125^2 + r^2} - 125\right)$$

8

In contrast, all C3-symmetry lenses are designed for linear focusing of light (m=1) at 0.8 µm wavelength at 80 µm from the surface, resulting in focusing at 110 µm for SHG focusing (m=2) at 0.4 µm.

$$\Theta = \frac{2\pi}{2\lambda_{\text{fund}}}\left(\sqrt{80^2 + r^2} - 80\right) = \frac{2\pi}{3(\lambda_{\text{fund}}/2)}\left(\sqrt{110^2 + r^2} - 110\right)$$

These focal distances used in the design of the lenses can be precisely reproduced by the measurements in Fig. SI 7.

## 10. Protocol for determining the SHG power density enhancement

The SHG intensity/power density enhancement of the metalens was determined by comparing the speckle patterns of the metasurface and the metalens focal spot as described in Figure SI 9. The analysis area for the surface was determined by measuring the full-width-half-maximum (FWHM) of the pump beam intensity squared (Fig. SI 9 a). Equivalent excitation area was considered when analyzing the SHG emission from the metalens surface (Fig. SI 9 b and c) and average intensity was determined by summing all pixel intensities and dividing by the total number of pixels in the circular region of interest (Fig. SI 9 d).

Same steps were followed for determining the average pixel intensity at the focal spot (Fig. SI 9 e-h). Acquired intermediate results are presented in Table SI 2.

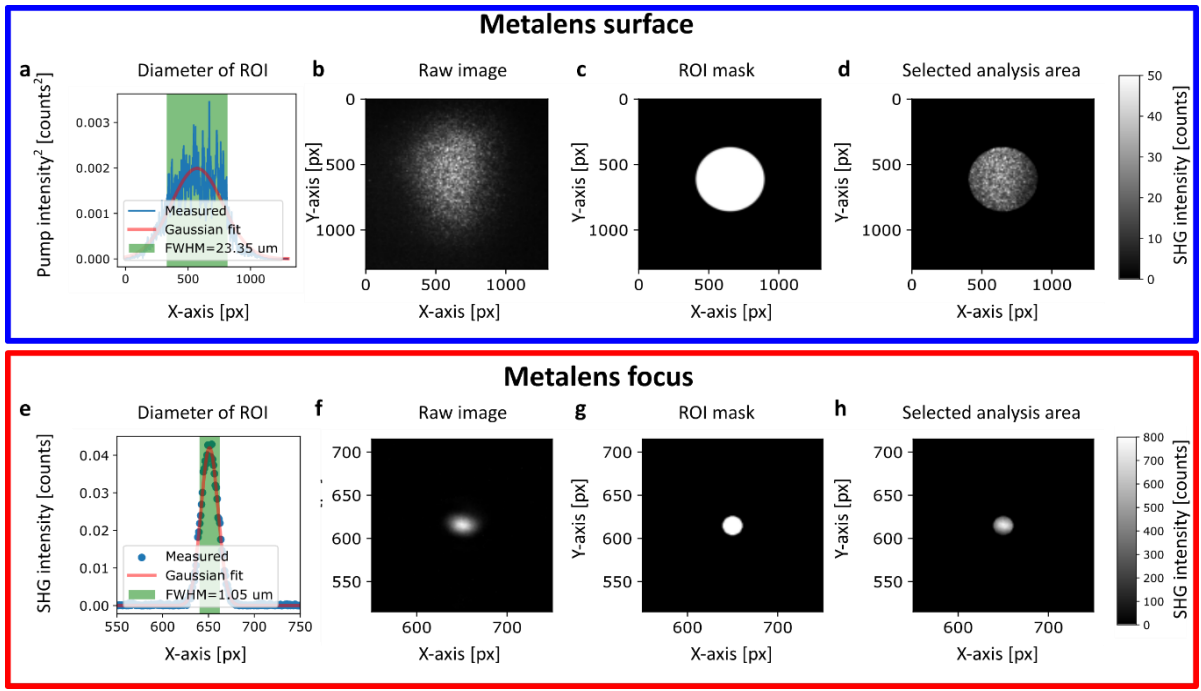

*Figure SI 9 Example of image analysis steps to determine the SHG intensity enhancement by the metalens focus. a) Pump beam diameter at metalens surface determined by fitting pump intensity cross-section squared to a Gaussian curve. b) Acquired SHG image at metalens surface. c) Region of interest (ROI) mask with diameter determined in part a. d) ROI of metalens surface for average power density (counts/pixel) estimation. e) SHG beam diameter at metalens focus determined by fitting SHG intensity cross-section to a Gaussian curve. f) Acquired SHG image at metalens focus. g) ROI mask with diameter determined in part e. h) ROI of metalens focus for average power density (counts/pixel) estimation*

*Table SI 2. Broadband metalens characterization results obtained using the protocol described in Figure SI 5.*

| Pump wavelength [nm] | Total intensity surface [counts] | Total intensity focus [counts] | FWHM focus [pixels] | FWHM focus [um] | Focusing efficiency [%] | SHG power density enhancement |
|---|---|---|---|---|---|---|
| 760 | 1169615 | 28485 | 21 | 1.0 | 2.4 | 12 |
| 775 | 2283845 | 71447 | 21 | 1.0 | 3.1 | 16 |
| 800 | 3761863 | 175749 | 22 | 1.1 | 4.7 | 23 |
| 825 | 5189136 | 283062 | 21 | 1.0 | 5.5 | 27 |
| 850 | 5324366 | 340220 | 22 | 1.1 | 6.4 | 31 |
| 875 | 2870543 | 156118 | 22 | 1.1 | 5.4 | 27 |
| 900 | 1843363 | 126443 | 21 | 1.0 | 6.9 | 34 |
| 950 | 1226621 | 51737 | 19 | 0.9 | 4.2 | 25 |
| 1550* | 436812 | 14018 | 42 | 2.0 | 3.2 | 2.6 |

*\*Different laser source, optical components and acquisition parameters were used as described in Figure SI 2.*


\end{document}